\documentclass[aps, notitlepage, floatfix, pre, superscriptaddress, twocolumn]{revtex4-2}
\usepackage[colorlinks=true, allcolors=blue, bookmarks=false]{hyperref}
\usepackage{amsfonts, amsmath, amssymb}
\usepackage{graphicx}

\graphicspath{{./figures/}}
\newcommand\mean[1]{\ensuremath{\left\langle#1\right\rangle}}

\begin{document}

\author{Wouter Buijsman}
\email{buijsman@pks.mpg.de}
\affiliation{Max Planck Institute for the Physics of Complex Systems, 01187 Dresden, Germany}

\author{Masudul Haque}
\email{masudul.haque@tu-dresden.de}
\affiliation{Institut f\"ur Theoretische Physik, Technische Universit\"at Dresden, 01062 Dresden, Germany}
\affiliation{Max Planck Institute for the Physics of Complex Systems, 01187 Dresden, Germany}

\author{Ivan M. Khaymovich}
\email{ivan.khaymovich@gmail.com}
\affiliation{Nordita, Stockholm University \& KTH Royal Institute of Technology, SE-106 91 Stockholm, Sweden}

\title{Power-law banded random matrix ensemble as a \\ model for quantum many-body Hamiltonians}

\date{\today}

\begin{abstract}
We explore interpretations of the power-law banded random matrix (PLBRM) ensemble as Hamiltonians of one-dimensional quantum many-body systems.  We introduce and compare a number of labeling schemes for assigning random matrix basis indices to many-body basis vectors.  We compare the physical properties of the resulting Hamiltonians, focusing on the half-system eigenstate bipartite entanglement entropy. We show and quantify how the different PLBRM phases (ergodic, weakly ergodic, localized), known from the single-particle interpretation, can be interpreted as entanglement transitions in the quantum many-body interpretation.  For the weakly ergodic phase, where spectral edge and bulk eigenstates show distinct behavior, we perform a detailed scaling analysis to provide a quantitative picture of the boundaries between different types of entanglement scaling behaviors. In particular, we identify and characterize an intermediate set of eigenstates whose entanglement entropy have volume law scaling but nonvanishing deviation from the Page value expected for maximally ergodic states.

\end{abstract}

\maketitle

\section{Introduction}

Ideas from random matrix theory~\cite{Mehta91, Guhr98, Haake10} are central to our current understanding of isolated quantum systems and quantum many-body chaos~\cite{Deutsch_PRA1991, Srednicki_PRE1994, Rigol_Santos_PRA2010, Reimann_PRL2015, Dalessio2016, Borgonovi16, Gogolin_Eisert_RepProgPhys2016,  Mondaini_Rigol_PRE2017, Deutsch_RepProgPhys2018, KhaymovichHaqueMcClarty_PRL2019, Sugimoto_Hamazaki_Ueda_PRL2021}.  Away from the edges of the many-body spectrum, important aspects of physical eigenstates and eigenvalues are well-described by random matrix models, provided the system is not integrable or many-body localized.  
The most prominent random matrix ensembles used to model quantum Hamiltonians are the Gaussian orthogonal (GOE) and unitary (GUE) ensembles, for which the matrix elements are uncorrelated and identically distributed.

However, physical Hamiltonians expressed in a natural local basis often exhibit a structure: basis states $|n \rangle$ and $|m \rangle$ tend to be stonger (weaker) coupled when $|n - m|$ is small (large).
Several random matrix models or their variants incorporate this idea, e.g., the Rosenzweig-Porter model~\cite{Rosenzweig60, Kravtsov15, Facoetti16, Truong16, Amini17, vonSoosten17, Monthus2017multifractality} and recently proposed variations thereof~\cite{Kravtsov20, Khaymovich20, Biroli21, Khaymovich21, Buijsman22, DeTomasi22, Venturelli23, Sarkar23, Kutlin24, Buijsman_PRB2024_RP, Ghosh25}, the $\beta$ ensemble~\cite{DumitriuEdelman_JMathPhys2002_beta, Buijsman19, Balasubramanian_Magan_arxiv2023_tridiagonalizing, Das_Ghosh_PRE2022_beta,  DasGhoshKhaymovich_PRL2023_absence, DasGhoshKhaymovich_PRB2024_beta}, banded random matrices~\cite{Wigner_1955_banded, Casati_Molinari_Izrailev_PRL1990_band, Fyodorov_Mirlin_PRL1991_banded, Deutsch_PRA1991, Fyodorov_Mirlin_PRL1992_banded, Fyodorov_Mirlin_PRL1993_banded, ProsenRobnik_JPhysA1993_sparsebanded, Frahm_MuellerGoerling_EPL1995_banded, Casati_Chirikov_Izrailev_PhysLettA1996_banded, Fyodorov_Izrailev_Casati_PRL1996_banded, Deutsch_RepProgPhys2018, DasGhoshSantos_2025_SFF_banded}, 
ultrametric random matrix models~\cite{Fyodorov09, Bogomolny11, Rushkin11, Gutkin11, MendezBermudez12, vonSoosten18, Bogomolny18, vonSoosten19, Suntajs_Hopjan_DeRoeck_Vidmar_PRR2024_avalanche_ultrametric}, and power-law banded random matrix (PLBRM) models~\cite{Mirlin96, Kravtsov_Muttalib_PRL97, Varga00, Mirlin_Evers_PRB2000, Cuevas_Kravtsov_PRB2007,  Evers08, Bogomolny11, Rushkin11, MendezBermudez12, Bogomolny18, DeTomasi19, Nosov2019correlation, VegaOliveros19, Lydzba_Rigol_Vidmar_PRL2020_eigenstateEE_randomquadratic, Haque22, Rao22, Hopjan_Vidmar_PRR2023_ScaleInvariant, Lydzba_Mierzejewski_Rigol_Vidmar_PRB2024_NormalWeakEigThermalizn, Santra25}.  The last is the focus of the present work.  

Power-law banded random matrices have uncorrelated (up to Hermiticity) normally distributed elements with mean zero and a variance that decays in a power-law fashion as a function of the distance from the main diagonal.  Determined by the power law exponent, the eigenvectors can be fully or weakly ergodic, multifractal, or localized. Interpreting PLBRMs as Hamiltonians of one-dimensional single-particle quantum systems has proven to be a meaningful exercise in studying Anderson localization~\cite{Mirlin_PhysRep2000, Evers00, Mirlin_Evers_PRB2000, Evers08, Quinto16, Hopjan_Vidmar_PRR2023_ScaleInvariant}.  Recently, generalized forms of PLBRMs have been used to study ergodicity breaking in non-Hermitian~\cite{DeTomasi23, Ghosh23, VallejoFabila24, Prasad_Tekur_Kulkarni_PRB2025_nonHerm_PLBRM} and periodically driven~\cite{Tiwari24} systems. 

In this work we explore many-body interpretations of the power-law banded random matrix ensemble. {e interpret PLBRMs as Hamiltonians of spin-$\frac{1}{2}$ systems not possessing any conservation laws. This means that, for an $L$-site system, the matrix dimension $N$ is $2^L$. The usual random-matrix  description of many-body Hamiltonians with matrix classes GOE or GUE has the shortcoming that the eigenvectors of such random matrices have similar properties in the spectral bulk and at the spectral edges.  In contrast, for physical many-body Hamiltonians, the low-energy eigenstates have markedly different properties compared to the eigenstates in the spectral bulk. Perhaps the best known manifestation of this spectral edge-bulk distinction is seen in the entanglement entropy ($S_\text{ent}$) of the eigenstates: At the spectral edges, $S_\text{ent}$ is low (``area law'') \cite{Eisert_Plenio_RMP2010_AreaLaws, Amico_Fazio_RMP2010_EntanglementReview}, while in the mid-spectrum ``infinite-temperature'' regime, the eigenstates have $S_\text{ent}$ close to the value expected for random states, the so-called Page value $S_\text{Page}$  (``volume law'').  As a result, for chaotic many-body systems, a scatter plot of $S_\text{ent}$ versus eigenenergy takes the shape of an arch or rainbow, by now familiar from many numerical examples \cite{Beugeling_Andreanov_Haque_JSM2015, Garrison_Grover_PRX2018_SingleEigenstate, Turner_Abanin_Serbyn_scarred_PRB2018,  Moudgalya_Regnault_Bernevig_AKLT_entngl_PRB2018, Mudry_Castelnovo_Chamon_Neupert_PRRes2019,  Sen_Sen_Sengupta_Floquetscars_PRB2020, Sen_Sen_Sengupta_drivenRydberg_2020,  Iadecola_Schecter_PRB2020, Mark_Lin_Motrunich_PRB2020, Shibata_Katsura_OnsagerScars_PRL2020,  Pichler_Lukin_Ho_PRB2020_xy, Mark_Motrunich_2020_etapairing, McClarty_Haque_Sen_Richter_2020, Haque22, Bianchi_Hackl_Kieburg_Rigol_Vidmar_PRXQ2022, Tamura_Katsura_PRB2022_scars, Kliczkowski_Vidmar_Rigol_PRE2023, Swietek_Vidmar_Rigol_PRE2024_entngmnt_integrableXYZ, Imai_Tsuji_PRRes2025_scars}. The absence of this structure in GOE or GUE random matrices is rather striking, and is a motivation to study the many-body interpretation of other, more structured, random matrix ensembles.  In particular, it has been noted~\cite{Haque22} that power-law banded random matrices reflect this distinction between edge and bulk eigenstates in the weakly ergodic phase.  
Thus, this random matrix ensemble provides an improved model of the structure of chaotic many-body Hamiltonians. This background also provides motivation to study in detail the entanglement entropy of the eigenstates of PLBRMs with a many-body interpretation.  This is a main focus of the present work.  

When interpreting GOE or GUE matrices as physical Hamiltonians, the basis labels can be assigned arbitrarily to the physical basis states, since these matrices have no structure.    For interpreting structured random matrices as Hamiltonians of quantum many-body systems,  assigning a physical interpretation to the basis vectors is a trickier issue.  One wants to do this in such a way that the resulting Hamiltonian resembles or models a system with only few-body, short-range interactions, which are the hallmarks of physical Hamiltonians.  In Refs.~\cite{Haque22, Rao22}, each basis vector is assigned a (different) many-body product state, where the index is obtained by interpreting it as a binary number of $L$ bits, where $L$ is the number of lattice sites.

In the present work, we provide a more systematic study of interpreting PLBRMs as Hamiltonians of quantum many-body systems. After a brief review of PLBRMs (Sec.~\ref{sec: PLBRM}), we first identify three different ways to label basis vectors with many-body configurations (Sec.~\ref{sec: labeling}). We quantify the quality of the labeling schemes, and argue one of our labeling schemes to improve on the previously employed binary scheme.  We show how to resolve an issue regarding the homogeneity of the resulting Hamiltonians using what we will refer to as ``site randomization''.  Next (Sec.~\ref{sec: entanglement}), we turn to the physical properties of the resulting Hamiltonians, focusing on the entanglement entropy.  We study the scaling of the eigenstate entanglement entropy with system size in the different phases of the model.

The general picture is that $S_\text{ent}$ exhibits volume law behavior throughout the spectrum in the fully ergodic phase ($\alpha<\frac{1}{2}$), area law behavior throughout the spectrum in the localized phase ($\alpha>1$), while in the intermediate weakly ergodic phase  ($\frac{1}{2}<\alpha<1$) there is a bulk-edge difference with the bulk showing volume law and the edge showing area law.  We confirm this picture quantitatively through separate finite-size scaling analyses for bulk $S_\text{ent}$ and edge $S_\text{ent}$ (Sec.~\ref{sec: transitions}).  The weakly ergodic phase shows the most interesting variation between the different parts of the spectrum, and this is also the behavior familiar from many-body systems.  Therefore we undertake a more detailed study of this phase, seeking to identify the demarcation between bulk and edge eigenstates (Sec.~\ref{sec: boundary}).  Using scaling analysis on different quantities, we numerically identify an intermediate set of eigenstates, which have volume law scaling but a nonvanishing deviation from the Page value.  There are thus \emph{two} boundaries in energy.  One boundary separates eigenstates whose entanglement entropy approaches the Page value ($S_\text{ent}{\to}S_\text{Page}$) from these intermediate eigenstates. 
A second boundary separates these intermediate eigenstates from the area law eigenstates at the very spectral edge.  We conclude (Sec.~\ref{sec: conclusions}) with a  summary and discussion and some questions opened up by this work for future investigations.

\section{Power-law banded random matrices} \label{sec: PLBRM}
 Various related definitions of the power-law banded random matrix ensemble are in use in the literature~\cite{Evers08}. Matrices $H$ sampled from such an ensemble are typically element-wise defined as
 \begin{equation}
 H_{ij} = G_{ij} \, a(| i -j|),
\end{equation}
where $G_{ij} = G_{ji}$ are independent and identically distributed random numbers, representing the GOE random matrix ensemble. The GOE consists of real-valued symmetric matrices with entries sampled independently from the Gaussian distribution with mean $\mu = 0$ and off-diagonal (diagonal) components with variance $\sigma^2 = 1 / 2$ ($\sigma^2 = 1$)~\cite{Mehta91}.  The function $a(r)$ decays as a power law  for $r \gg 1$, with a tunable exponent $\alpha > 0$. We will use
\begin{equation}
a(r) = \frac{1}{1 + (r / \beta)^\alpha},
\label{eq:a}
\end{equation}
with the so-called bandwidth $\beta$ set to unity ($\beta = 1$). This definition has been part of investigations previously in Ref.~\cite{Haque22}, involving some of the present authors. Other often used definitions can be found, for example, in Refs.~\cite{Mirlin96, Kravtsov_Muttalib_PRL97, Mirlin_Evers_PRB2000, Varga00, Cuevas_Kravtsov_PRB2007, Bogomolny11, Bogomolny18, Rao22, Hopjan_Vidmar_PRR2023_ScaleInvariant}.  
A few of these alternative forms are of the type $a(r)\sim 1/\sqrt{1 + (r/\beta)^{2\alpha}}$ or
$a(r)\sim 1/\left[1 + (r/\beta)^2\right]^{\alpha/2}$, or periodic versions thereof.  Like the form \eqref{eq:a} that we are using, these are also smooth functions satisfying the same limiting behaviors: $r\sim\text{const}$ for $r \ll \beta$ and $r\sim r^{-\alpha}$ for $r \gg \beta$. We expect our qualitative results to be independent of the explicit choice of $a(r)$, as long as the bandwidth $\beta$ is not large.

The parameter $\alpha$ can be used to tune the eigenstates in the bulk (middle) of the spectrum.  Depending on their behavior, three phases of the system have been identified: the fully ergodic  ($\alpha < \frac{1}{2}$)~\cite{vonSoosten18} phase, the weakly ergodic ($\frac{1}{2} < \alpha < 1$)~\cite{Bogomolny18} phase, and the (power-law) localized ($\alpha > 1$) phase~\cite{Mirlin96}. This $\alpha$-dependence is illustrated in Fig.~\ref{fig: phasediagram}.

In the fully ergodic phase ($\alpha<\frac{1}{2}$), the properties of the model are statistically the same as those of the GOE. This means that for an eigenstate $| \psi \rangle$, the inverse participation ratio $\text{IPR} = \sum_n | \langle n | \psi \rangle|^4$, with the summation running over all basis states $| n \rangle$, is asymptotically given by $3/N$, as is the case with fully ergodic states \cite{Mehta91}. Here $N$ is the matrix dimension ($n = 1, 2, \dots, N$). 

The weakly ergodic phase, $\alpha\in\left(\frac{1}{2},1\right)$, is characterized by bulk eigenstates occupying only a finite fraction of the Hilbert space~\cite{Bogomolny18, Baecker19}.  The inverse participation ratio of the eigenstates is asymptotically given by  $c/N$ for some $c > 3$ depending on $\alpha$.  

The fractal dimensions $D_q$ ($q \ge 1$) of an eigenstate $| \psi \rangle$ are defined through the scaling
\begin{equation}\label{eq:IPR}
\sum_n |\langle n | \psi \rangle|^{2q} \sim N^{-D_q (q-1)} . 
\end{equation}
The fully and weakly ergodic phases are characterized by eigenstates with unit fractal dimensions ($D_q = 1$). At the critical point $\alpha = 1$, the eigenstates in the bulk of the spectrum are multifractal, i.e., $D_q\in(0,1)$ is a $q$-dependent function, and level statistics are intermediate between Poisson and Wigner-Dyson. 

In the localized phase ($\alpha>1$), the eigenstates have fractal dimension zero for positive integers $q$, meaning that the inverse participation ratio does not scale as $1/N$ and the eigenstates only occupy a finite number of Hilbert-space configurations. Fig.~\ref{fig: phasediagram} provides a graphical summary of the phase diagram discussed here.

\begin{figure}[t]
\includegraphics[width = \columnwidth]{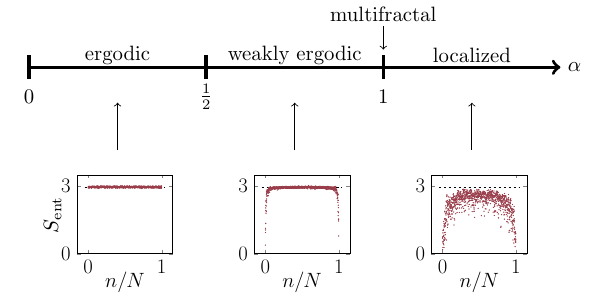}
\caption{Phase diagram of the power-law banded random matrix ensemble, showing the phases of mid-spectrum eigenstates as a function of the power-law exponent $\alpha$. The panels show the eigenstate entanglement entropy for a single realization as a function of the normalized eigenstate index (random labeling scheme, $L=10$), for the three phases ($\alpha = 0.25$, $\alpha = 0.75$, $\alpha = 1.25$). In the panels, the ground state is on the left of the horizontal axis, while the highest excited eigenstate is on the right. The horizontal dashed line indicates the Page  entanglement entropy $S_\text{Page}$. (Ensemble-averaged versions of such plots are discussed later, see Fig.~\ref{fig: full}.)}
\label{fig: phasediagram}
\end{figure}

The phenomenology of power-law banded random matrices can be understood from resonance counting~\cite{Levitov1989, Levitov1990} and the Breit-Wigner approximation (see, for example, Refs.~\cite{BogomolnyRP2018,Monthus2017multifractality}). First, from resonance counting arguments, it follows that localization perturbation theory converges if the number $N_\text{res}$ of resonances, defined as $|H_{mn}|>|H_{nn}-H_{mm}|$, does not grow with the size $N$ of the matrix~\cite{Khaymovich20, Khaymovich21, Nosov2019correlation}. To calculate $N_\text{res}$, one should sum the probabilities of the other nodes $m \neq n$ to be in resonance with node $n$. For independent diagonal elements $H_{ii}$ sampled identically and uniformly from $[-W/2, W/2]$, one obtains
\begin{equation}
N_\text{res} = \sum_{\substack{m=1 \\ m\ne n}}^N \int_{-W}^W P(H_{mn}) |H_{mn}| dH_{mn}.
\end{equation}
In the case of a normal distribution of $H_{ii}$ with $W$ related to the standard deviation, the above expression is the same in the leading order up to an unimportant prefactor.
For power-law banded random matrices, the above integral gives $\int_{-W}^W P(H_{mn}) |H_{mn}| dH_{mn} \sim R^{-\alpha}$ with $R = |n - m|$ and leads to $N_\text{res}\sim \sum_{R} R^{-\alpha}\sim N^{1-\alpha} + \mathcal{O}(1)$, which converges for $\alpha > 1$. Here and further, the summation runs over all admissible values of $R$, of which there are $\mathcal{O}(N)$. Therefore, for $\alpha > 1$ one expects localized eigenstates. 

Second, the Breit-Wigner approximation is based on Fermi's golden rule where the level broadening 
\begin{equation}
\Gamma_n(E) = \frac{2\pi }{\hbar} \nu(E) \sum_{\substack{m = 1 \\ m \ne n}}^N |H_{mn}|^2,
\end{equation}
with the normalized density of states $\nu(E)$, enters the wave-function estimate as
\begin{equation}
|\psi_E(n)|^2 \simeq \frac{A \Gamma_n(E)}{(E-H_{nn})^2 +\Gamma_n(E)^2},
\end{equation}
with a certain normalization prefactor $A$~\cite{BogomolnyRP2018,Monthus2017multifractality}. Here, $\psi_E(n)$ is the overlap of an eigenstate having eigenenergy $E$ with the basis state $|n \rangle$.
The broadening $\Gamma_n(E)$ in general depends both on the eigenvalue $E$ and on the location of the wave-function maximum $n$, where the time-dependent perturbation theory has been started.
As soon as $\Gamma(E)\equiv \mean{\Gamma_n(E)}\gg \mean{H_{nn}^2}$ for all the energies $E$, all the  wave-function coefficients are of the same order and, therefore, all the states should be ergodic. For power-law banded random matrices, one obtains the Breit-Wigner broadening to be $\Gamma_n(E) \sim \sum_R R^{-2\alpha}\sim N^{1-2\alpha} + \mathcal{O}(1)$. This diverges at $\alpha < \frac{1}{2}$ for all the energies. Therefore, for $\alpha < \frac{1}{2}$, one expects all the eigenstates to be ergodic (statistically equivalent to those of the GOE). Together, these two lines of arguments predict the `phase diagram' shown in Fig.~\ref{fig: phasediagram}. 

Regarding the intermediate region $\frac{1}{2} < \alpha < 1$ in-between the ergodic ($\alpha < \frac{1}{2}$) and localized ($\alpha > 1$) phases, it has been established that mid-spectrum eigenstates are weakly ergodic~\cite{Mirlin96,Bogomolny18}. More recent numerical results suggests that the spectral edge eigenstates behave differently~\cite{Haque22, Rao22}.   This difference between edge and bulk eigenstates is a strong reason for considering power-law banded random matrices to be an improved model for many-body Hamiltonians, as compared to non-banded matrices (GOE or GUE) which do not show such a difference. In this paper we will explore the difference between spectral bulk and edge eigenstates in detail.

In the quantum many-body interpretation, to be developed in the following sections, the two transitions appear to correspond to entanglement phase transitions \cite{Haque22}.   As one increases $\alpha$, we start off with all eigenstates having volume law behavior for $\alpha<\frac{1}{2}$.  As $\alpha$ crosses $\alpha=\frac{1}{2}$,  the edge eigenstates change in nature and acquire area law scaling, and then when $\alpha$ crosses $\alpha=1$,  the bulk eigenstates also turn to area law scaling.  These notions will be confirmed through numerical scaling analyses later on, but we have provided a visualization of the different behaviors in the three panels in Fig.~\ref{fig: phasediagram}.  The panels show single-realization scatter plots of $S_\text{ent}$ against eigenvalue index, using what we call the random labeling scheme (Sec.~\ref{sec: labeling}).

\section{Basis state labeling} \label{sec: labeling}

We aim to interpret the PLBRM ensemble as Hamiltonians of quantum many-body systems. This can be accomplished by associating each basis vector with a (different) many-body configuration.  We will focus on spin-$\frac{1}{2}$ chains in this work.  For simplicity, we restrict to systems without any conservation laws. The Hamiltonian of an $L$-site system is then modeled as a random matrix of dimension $N = 2^L$, and the many-body configurations (basis states) are strings of $L$ bits. Assigning many-body configurations to basis indices of the random matrix can be done in various ways.  We  propose and examine three different labeling schemes, which we refer to as `random', binary', and `Gray code'. The binary labeling scheme has been considered before in Refs.~\cite{Haque22, Rao22}. We devise a way to quantify the quality of a labeling scheme, and find the Gray code scheme to be of higher quality than the binary one. We find the (ensemble-averaged) Hamiltonians for the binary and Gray code labeling schemes to be non-homogeneous, meaning that the physical properties are dependent on the site index. We next discuss how to correct for this through what we will refer to as `site randomization'.

The issue of labeling scheme is significant when using banded random matrices as a model for many-body Hamiltonians.  This is in contrast to the case of using  random matrix models characterized by identically distributed off-diagonal elements (e.g., GOE or GUE).  In the latter  (more conventional) case, any labeling scheme would lead to the same Hamiltonian on average.

\subsection{Labeling schemes}

We represent many-body configurations $c_i$ (for example, $c_i = 001111$) with $i$ ranging over integers from $1$ to $N$ by $L$ binary digits, where the $j$th digit is zero (one) if the $j$th spin is in the down (up) state.
In power-law banded random matrices, the matrix elements between configurations with neighboring labels have larger magnitude on average.  Since physical Hamiltonians contain few-body operators, we thus want neighboring configurations to differ from each other by as few spin flips as possible.  

In what follows, we quantify the quality of a given labeling scheme by the ``badness''
\begin{equation}
B = \frac{1}{N - 1} \sum_{i=1}^{N-1} d_2(c_i, c_{i+1}).
\label{eq: badness}
\end{equation}
Here, $d_2(c_i, c_j)$ represents the Hamming distance between the many-body configurations $c_i$ and $c_j$, i.e., the minimum number of spin flips connecting the two configurations.  Here the subscript $2$ indicates that the numbers are represented in binary digits. The badness penalizes for successive many-body configurations $c_i$ and $c_{i+1}$ to be separated by a large number of spin flips.  

\begin{figure}[t]
\includegraphics[width = \columnwidth]{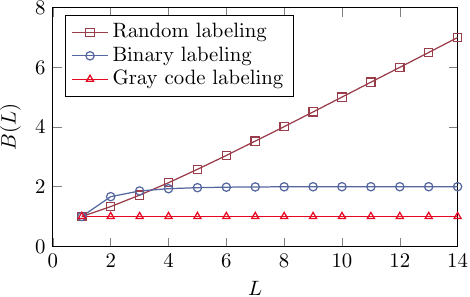}
\caption{Badness~\eqref{eq: badness} as a function of the system size $L$ for each of the labeling methods. For the random labeling scheme, the sample-to-sample variance is smaller than the marker size.}
\label{fig: badness}
\end{figure}

An obvious way to assign different many-body configurations to basis states is to assign randomly.  Interactions in the many-body picture of the random labeling scheme are non-local and tend to involve many spins. 

In the thermodynamic limit $L \gg 1$, the set of all possible many-body configurations is dominated by configurations with (close to) $L / 2$ up-spins. The average Hamming distance between two randomly selected basis states approaches then $L / 2$, since one (on average) needs to flip $L / 4$ spins from down to up, and $L / 4$ spins from up to down.  For the random labeling scheme, one thus has that $B(L) \approx L / 2$ for large $L$ (Figure \ref{fig: badness}). 

A lower badness can be achieved by adapting a labeling scheme that prefers the configurations $c_i$ and $c_{i+1}$ to be separated by a smaller number of spin flips. This can be accomplished by taking $c_i$ to be the binary representation of $i-1$. We refer to this as the `binary' labeling scheme; this was used in Ref.\ \cite{Haque22, Rao22}.  For $L = 3$, as an example, the labels are
\begin{align*}
c_1 = 000, \qquad & c_2 = 001, \qquad & c_3 = 010, \qquad & c_4 = 011, \\
c_5 = 100, \qquad & c_6 = 101, \qquad & c_7 = 110, \qquad & c_8 = 111.
\end{align*}
We find numerically (and it can be argued combinatorially) that the badness approaches $B(L) \approx 2$ for $L \gg 1$ for the binary labeling scheme (Figure \ref{fig: badness}).  This is significantly lower than the scaling $B(L) \approx L / 2$ for the random labeling scheme.

The Gray code provides an ordering of the binary numbers such that successive numbers differ by only a single element~\cite{Gray53, Doran07}. We refer to the corresponding labeling scheme as the ``Gray code'' labeling scheme, which is characterized by the lowest possible badness $B(L) = 1$. The configuration $c_n$ in the binary labeling scheme can be converted to the corresponding configuration in the Gray code labeling scheme by taking all up to the first nonzero digit identical. The subsequent digits are equal to $1$ if the corresponding and previous digit in the binary representation are different (i.e., $0$ and $1$), and $0$ otherwise. Taking $L = 3$ again as an example, one finds
\begin{align*}
c_1 = 000, \qquad & c_2 = 001, \qquad & c_3 = 011, \qquad & c_4 = 010, \\
c_5 = 110, \qquad & c_6 = 111, \qquad & c_7 = 101, \qquad & c_8 = 100.
\end{align*}
We note that there is no unique labeling scheme with the lowest possible badness (e.g., the badness does not change if the ordering of the digits is changed). Figure~\ref{fig: badness} shows the badness as a function of the system size for the random, binary, and Gray code labeling methods. Clearly, the asymptotic scalings are well-approximated already at small system sizes. 

The badness~\eqref{eq:  badness} used to quantify the quality of a chosen basis state labeling scheme is rather simple. In principle, it is possible to define more sophisticated probes that are sensitive to other features one might wish a labeling scheme to have. For example, one might want to penalize successive configurations for having pairs of spin flips which are far from each other, and one might want to consider pairs of configurations with labels differing by more than one. For the purposes of the present discussion, the badness as defined in Eq.~\eqref{eq: badness} is sufficient.

\subsection{Site randomization}
Despite having a low badness, the binary and Gray code labeling schemes have a feature that is unwanted in the present study: they do not model spatially homogeneous  Hamiltonians. For the binary labeling scheme, changing the $n$-th digit (counted from the right and starting at zero) from zero to one in the binary representation increases the index that is represented by $2^n$. For Hamiltonians given by power-law banded random matrices, such a hierarchy between basis states means that interactions between spins are weaker for sites on the left than for sites on the right.  

We illustrate this spatial asymmetry using the entanglement entropy of mid-spectrum eigenstates for the binary labeling scheme (Fig.~\ref{fig: asymmetry}).   The entanglement entropy of fully ergodic (structureless) eigenstates for a decomposition in subsystems $A$ (size $L_A$) and $B$ (size $L_B$) is well-approximated by the Page value~\cite{Lubkin78, Page93}
\begin{equation}
S_\text{Page} = \ln(2) \, \min(L_A, L_B) - \frac{2^{\min(L_A, L_B)}}{2^{\max(L_A, L_B) + 1}} \ .
\label{eq: SPage}
\end{equation}
Here, subsystems $A$ and $B$ cover the leftmost $L_A$ and rightmost $L_B = L - L_A$ sites, respectively. Fig.~\ref{fig: asymmetry} shows the ensemble-averaged eigenstate entanglement entropy $S_\text{ent}$ of mid-spectrum eigenstates  as a function of the subsystem fraction $L / L_A$ for various total system sizes.  The Page predictions~\eqref{eq: SPage} are shown as dotted lines for reference.  We use the exponent $\alpha=1$, for which the effect is prominent. 

\begin{figure}
\includegraphics[width = \columnwidth]{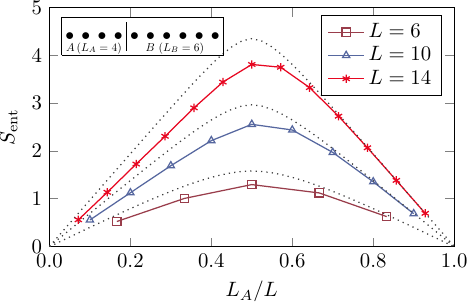}
\caption{Ensemble-averaged eigenstate entanglement entropy for mid-spectrum eigenstates with $\alpha = 1$, computed using the binary labeling scheme (no site randomization). The spin chain is decomposed in subsystems $A$ and $B$ consisting of the first $L_A$ and last $L_B = L - L_A$ sites, respectively. The inset illustrates the decomposition for $L_A = 4$ and $L_B = 6$. The dotted lines give the Page values~\eqref{eq: SPage} for $L=6$, $L=10$, and $L=14$.}
\label{fig: asymmetry}
\end{figure}

For $L_A / L \approx 1$, the eigenstate entanglement entropy is close to the Page value (dotted line). The entropy is much smaller than the Page value for $L_A / L \approx 0$. To understand this asymmetry, we start by noting that eigenstates that are not maximally ergodic are not uniformly spread out over all Hilbert space~\cite{Borgonovi16}. Such eigenstates tend to have the same binary digits near the left spatial edge (covering subsystem $A$), indicating that they are mainly supported in a specific `region' of the Hilbert space. On the contrary, a change in the digits near the right spatial edge (covering subsystem $B$) changes the resulting basis state number only slightly, meaning that the state is in the same Hilbert space region before and after the change in the digits. Hence, there is no tendency for these digits to be the same, leading to non-symmetric curves when plotting the eigenstate entanglement entropy as a function of $L_A / L$. This asymmetry is also seen for the Gray code labeling scheme, and the same argument can be used to explain the effect. We stress that the discussion here is about the \emph{spatial} edges, not the \emph{spectral} ones.

In what follows, every time we use binary labeling or Gray code labeling, we randomize the $L$ real-space site indices, corresponding to the binary digits of the many-body configurations, to avoid the (ensemble-averaged) Hamiltonians being nonhomogeneous. We refer to this procedure as ``site randomization''. For the ensemble-averaged results shown below, both the Hamiltonian matrix and the site ordering have been sampled independently for each realization.

\begin{figure}[t]
\includegraphics[width = \columnwidth]{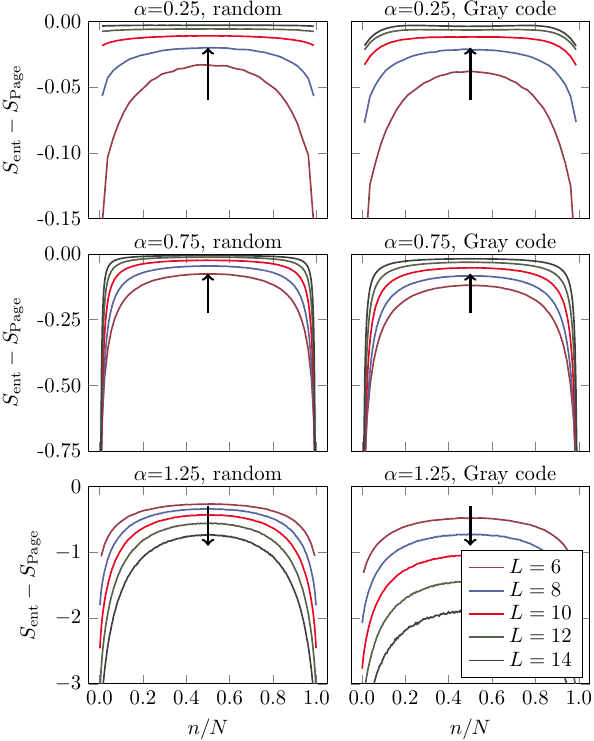}
\caption{Difference between the ensemble-averaged eigenstate entanglement entropy and the Page value as a function of the eigenstate index $n$ (starting from $n = 1$ and ordered by increasing energy), scaled by the Hilbert space dimension $N$. On the horizontal axes, the ground states are on the left, while the highest excited eigenstates are on the right. Left and right panels use random and Gray code labeling schemes respectively. The three rows correspond to the three phases ($\alpha = 0.25$, $\alpha = 0.75$, $\alpha = 1.25$). The arrows indicate the direction of flow for the bulk eigenstates with increasing system size. Note the very different scales on the vertical axes for the different values of $\alpha$.}
\label{fig: full}
\end{figure}

\section{Eigenstate Entanglement: Overall Features} \label{sec: entanglement}

We now analyze the many-body interpretations of the PLBRM ensemble. Our focus is on the ensemble-averaged eigenstate entanglement entropy for a bipartition of the system in left and right halves of equal size ($L_A=L_B = L/2$). Fig.~\ref{fig: full} illustrates its dependence on the power-law exponent $\alpha$, the basis state labeling scheme, the system size, and the location on the energy spectrum.  The left panels are ensemble-averaged versions of the single-realization scatter plots we presented in Fig.~\ref{fig: phasediagram}.  The eigenstate index $n$ starts from $n = 1$ and is ordered by increasing energy, meaning that the ground states are on the left, while the highest excited eigenstates are on the right of the horizontal axes.

In general, adopting the random labeling scheme (left panels) leads to values of $S_\text{ent}$ that are closer to $S_\text{Page}$ than those found using the Gray code labeling scheme (right panels). The eigenstates in the bulk (middle part) of the spectrum have different scaling behavior ($L$-dependence) for $\alpha<1$ and for $\alpha>1$, as shown using black arrows in Fig.~\ref{fig: full}.  In the fully and weakly ergodic phases (top row and center row), $S_\text{ent}$ gets closer to $S_\text{Page}$ with increasing system size.  In the localized phase (bottom row), $S_\text{ent}$ deviates away from $S_\text{Page}$ with increasing $L$. For the eigenstates near the spectral edges, $S_\text{ent}$ flows toward $S_\text{Page}$ in the fully ergodic phase and away from $S_\text{Page}$ in the localized phase, i.e., the direction is the same as the bulk (top and bottom rows of Fig.~\ref{fig: full}).  For the weakly ergodic phase (center row), the edge $S_\text{ent}$ moves away from $S_\text{Page}$ with increasing $L$, i.e., opposite to the arrow direction shown for the bulk $S_\text{ent}$.  This scaling is not obvious from Fig.~\ref{fig: full} but will be demonstrated in later figures.  

Thus, although all the curves in Fig.~\ref{fig: full} nominally have rainbow shapes, only the weakly ergodic phase $\alpha\in\left(\frac{1}{2},1\right)$ shows a true rainbow in the strong sense that the spectral bulk and spectral edge eigenstates have opposite scalings with system size. In the next two subsections we focus separately on the spectral bulk and edge eigenstates.

\subsection{Spectral bulk eigenstates}
Fig.~\ref{fig: bulk} shows the entanglement entropy and the difference from the Page value for mid-spectrum eigenstates as a function of $\alpha$ for the random and Gray code labeling schemes at several system sizes. For $\alpha \lesssim 1$, we observe volume-law ($S_\text{ent} \sim L$) scaling for both labeling schemes. In fact, the data suggests that $S_\text{ent} \approx S_\text{Page}$ in the thermodynamic limit. For $\alpha \gtrsim 1$, we observe area-law scaling ($S_\text{ent} \sim L^0$), again for both labeling schemes.  The random labeling scheme appears to be more sensitive to finite-size effects. The difference $S_\text{Page} - S_\text{ent}$ shows an approximate crossing for different system sizes at $\alpha \approx 1$, again for both labeling schemes. In Sec.~\ref{sec: transitions}, we study this crossing, which marks the transition between volume-law scaling and area-law scaling, quantitatively.

\begin{figure}[t]
\includegraphics[width = \columnwidth]{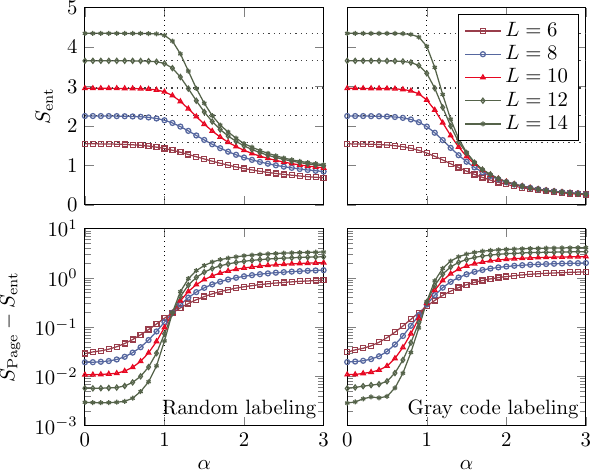}
\caption{Ensemble-averaged  entanglement entropies of mid-spectrum eigenstates, with indices $n$ ranging from $N / 2 - 10$ to $N / 2 + 10$.   Top row shows $S_\text{ent}$ and bottom row shows $S_\text{Page}-S_\text{ent}$. The left and right panels use the random and Gray code labeling scheme respectively. In the top row, the horizontal dashed lines give the Page values for the system sizes under consideration.}
\label{fig: bulk}
\end{figure}

\subsection{Spectral edge eigenstates}
Figure~\ref{fig: edge} shows the same plots as discussed above, but now for the spectral-edge eigenstates.  For small $\alpha$, we have volume-law ($S_\text{ent} \sim L$) scaling as we did for the bulk eigenstates, however, this behavior persists only up to $\alpha \approx \frac{1}{2}$. The crossing separating regimes with volume-law and area-law scaling of $S_\text{ent}$ is now located at $\alpha \approx \frac{1}{2}$.  (In contrast to $\alpha\approx1$ seen in Fig.~\ref{fig: bulk} for the bulk eigenstates.) We will study this crossing quantitatively as a function of system size in Section~\ref{sec: transitions}. 

The difference in behavior between bulk and edge eigenstates, particularly in the weakly ergodic phase $\frac{1}{2} < \alpha < 1$, raises the question of the boundary between edge and bulk.  This demarcation between edge and bulk, and how the demarcation varies with $\alpha$ in the weakly ergodic phase, will be analyzed in Section \ref{sec: boundary}.

\begin{figure}[t]
\includegraphics[width = \columnwidth]{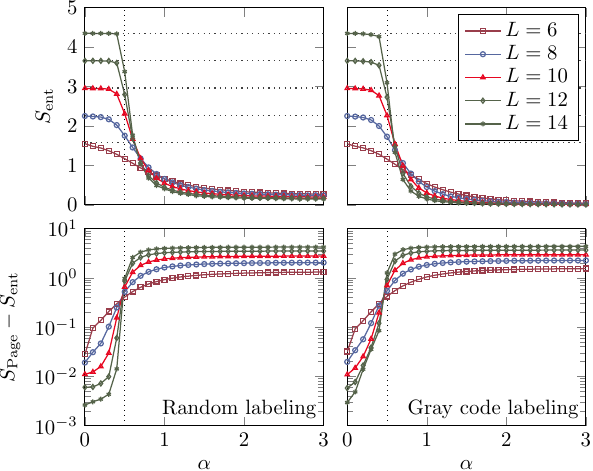}
\caption{Similar to Fig.~\ref{fig: bulk}, we show the ensemble-averaged entanglement entropy, but now for the eigenstates corresponding to the eigenvalues with the highest and lowest energies (that is, the ground state and the antiground state). The vertical dotted lines mark $\alpha = \frac{1}{2}$.}
\label{fig: edge}
\end{figure}

\section{Finite-size dependence} \label{sec: transitions}

In Figures \ref{fig: bulk} and \ref{fig: edge}, we saw crossings near $\alpha\approx1$ and near $\alpha\approx \frac{1}{2}$ for the spectral bulk and spectral edge eigenstates, respectively.  In this section, we present a finite size scaling analysis to determine the value $\alpha_0$ of the power-law exponent $\alpha$ at which the scaling of the entanglement entropy changes from volume-law to area-law in the thermodynamic limit.  Based on the theory reviewed in Sec.\ \ref{sec: PLBRM}, and based on the data in Figs.~\ref{fig: bulk} and~\ref{fig: edge}, we expect $\alpha _0= 1$ for the spectral bulk and $\alpha_0 = \frac{1}{2}$ for the spectral-edge eigenstates.

The quantity $\alpha_0$ can be estimated from the numerical data through various finite-size scaling procedures.  We have performed several scaling analyses, leading to consistent results.  Here we present one such analysis. 
We quantify the value $\alpha_0$ as the point where
\begin{equation}
S_\text{ent} = S_\text{Page} - \gamma,
\label{eq: condition}
\end{equation}
after ensemble-averaging, for some reasonable value of $\gamma$.  We will investigate this for several values of $\gamma > 0$. Figs.~\ref{fig: bulk} and~\ref{fig: edge} show that this point can be found for all system sizes roughly over the range $\gamma \in [0.05, 0.5]$. 
Empirically, we find that plotting the corresponding values as a function of $1/L^2$ allows for an extrapolation towards the thermodynamic limit in all settings that we consider. 

\begin{figure}[t]
\includegraphics[width = \columnwidth]{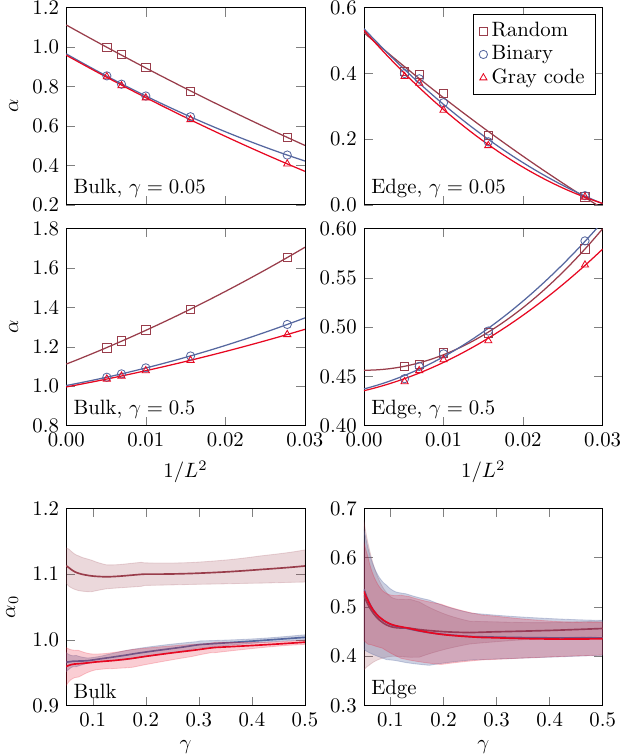}
\caption{Values of $\alpha$ for which $S_\text{Page} - S_\text{ent} = \gamma$, as a function of $1/L^2$ (after ensemble-averaging). The squares, circles, and triangles denote data for the random, binary, and Gray code labeling scheme, respectively. The spectral bulk (left column) and edge (right column) eigenstates are the same as the ones considered in Figs.~\ref{fig: bulk} and~\ref{fig: edge}, respectively. The solid lines give least-square fits of the form~\eqref{eq: fit}. The lower panels show the extrapolated value $\alpha_0$ for $L \to \infty$ as a function of $\gamma$. The shaded regions in the lower panels indicate the $95\%$ confidence interval of the fit~\eqref{eq: fit}.}
\label{fig: transition}
\end{figure}

Figure~\ref{fig: transition} shows the values of $\alpha$ at which condition~\eqref{eq: condition} holds as a function of $1 / L^2$ for both the spectral bulk (left panels) and edge (right panels) eigenstates for $\gamma = 0.05$ (top panels) and $\gamma = 0.5$ (middle panels). The connecting lines are least-square fits of the second-order polynomial
\begin{equation}
\alpha = \alpha_0 + \alpha_1 (1 / L^2) + \alpha_2 (1 / L^2)^2,
\label{eq: fit}
\end{equation}
with (least-squares) fitting parameters $\alpha_0$, $\alpha_1$, and $\alpha_2$. We observe that these fitted lines connect all data points reasonably well, although the fitting quality is better for the spectral bulk than for the spectral edge eigenstates. We focus on generic values of $\gamma \in [0.05, 0.5]$ in the bottom panels, which show the estimated transition point $\alpha_0$ as a function of $\gamma$ together with their uncertainty [$95\%$ confidence interval of the fit~\eqref{eq: fit}]. The estimates of $\alpha_0$ for the binary and Gray code labeling schemes are near their expected values over the full range. We observe a significant deviation for the random labeling scheme.  This means that the entanglement properties of the PLBRM ensemble interpreted as Hamiltonians of quantum many-body systems depend on the choice of the labeling scheme. For small values of $\gamma$, the condition is met for values of $\alpha$ below the crossing, leading to a flow towards larger values with increasing system size. The opposite occurs for large values of $\gamma$.

\begin{figure}[t]
\includegraphics[width = \columnwidth]{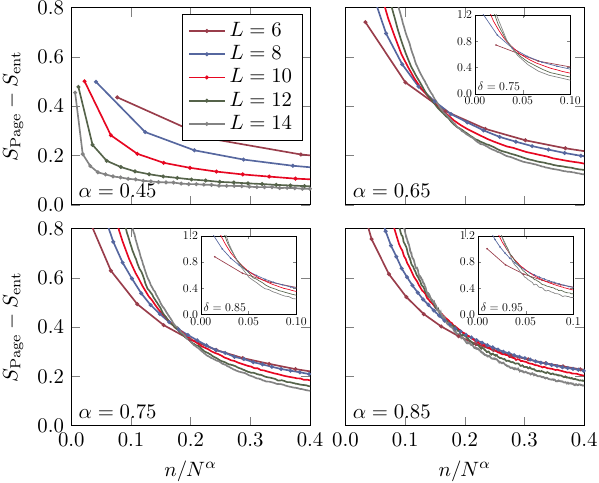}
\caption{Difference between the ensemble-averaged eigenstate entanglement entropy and the Page value for the Gray code labeling scheme as a function of the eigenstate index $n$, scaled by $N^\alpha$. The insets show the same quantity as a function of $n / N^\delta$ for $\delta = \alpha + 0.1$. For clarity, markers are shown only when the horizontal distance between data points is non-negligible.}
\label{fig: intersection}
\end{figure}

\section{Boundary between bulk and edge} \label{sec: boundary}

A natural follow-up question is how the transitions at $\alpha \approx \frac{1}{2}$ for the spectral edge and $\alpha \approx 1$ for the spectral bulk eigenstates are related to each other.  This relates to the question of how the spectrum is partitioned between bulk and edge parts, at various $\alpha$ in the weakly ergodic phase.  We identify two boundaries between different entanglement behaviors in the PLBRM spectrum.   In this section, we will restrict to the Gray code labeling scheme. 

In the weakly ergodic phase $\frac{1}{2} < \alpha < 1$, the spectral bulk eigenstates show volume law entanglement, while the spectral edge eigenstates show area law entanglement. 
The spectral bulk eigenstates form a measure one fraction of all eigenstates, while the number of spectral edge eigenstates is subextensive. Naturally, one expects the number of edge eigenstates (i) to increase with $N$, and (ii) to become a zero fraction in the thermodynamic limit. Although various scaling forms obey these properties, $N^\delta$ with $\delta < 1$ is arguably the simplest form. Therefore, we hypothesize that $N^\delta$ eigenstates at the spectral edges show non-bulk behavior, where $\delta$ is a nontrivial exponent.  Depending on the types of entanglement behaviors we want to differentiate, we introduce below two separate exponents, which we call $\delta_1$ and $\delta_2$.   

Normalizing the eigenstate index $n$ by $N^{\delta_1}$, we looked for a finite-size crossing of curves obtained by plotting $S_\text{Page} - S_\text{ent}$ as a function of $n/N^{\delta_1}$.    We find surprisingly good results for $\delta_1 = \alpha$ (Fig.~\ref{fig: intersection}).  For any fixed $\alpha$ within the weakly ergodic phase, $\frac{1}{2} < \alpha < 1$, we see that the curves for the different system sizes cross at fixed $n/N^\alpha \approx 0.2$.  The difference $S_\text{Page}-S_\text{ent}$ increases with system size below this energy and decreases with system size above this energy.  When $n$ is scaled by $N^{\delta}$ with $\delta \neq \alpha$, the curves no longer cross at approximately the same point, as shown in the insets for $\delta = \alpha + 0.1$. The nontrivial exponent $\delta_1$ which separates the two behaviors is thus (at least approximately) given by $\delta_1=\alpha$. At present we are not aware of an analytic derivation of this result.

Fig.~\ref{fig: critical} elaborates further on this by showing the ensemble-averaged eigenstate entanglement entropy for the $0.2 N^\alpha$ eigenstates with the lowest and highest  energies as a function of $\alpha$ for several system sizes.  
The quantity $S_\text{Page} - S_\text{ent}$ (left panel) shows a crossing at $\alpha \approx 0.5$, while $S_\text{ent} / S_\text{Page}$ (right panel) shows a crossing at $\alpha \approx 1$.  
In the weakly ergodic phase $\frac{1}{2} < \alpha < 1$, the deviation  $S_\text{Page} - S_\text{ent}$  increases with system size, but the ratio $S_\text{ent} / S_\text{Page}$ is also non-decreasing.  
This implies that, in the weakly ergodic phase, the entanglement entropy of the $\approx 0.2N^\alpha$ eigenstates at the spectral edges have nonvanishing deviation from $S_\text{Page}$, but the majority of these states still have volume-law scaling. 

Because $\delta_1=\alpha$, when we approach the fully ergodic phase, $\alpha\to \frac{1}{2}+0^+$, the number of eigenstates with nonvanishing deviation is $\sim N^{1/2}$.  This is a diverging number in the thermodynamic limit.  In the fully ergodic phase, $\alpha=\frac{1}{2}+0^-$, this number vanishes as all eigenstates have  $S_\text{ent}\to S_\text{Page}$ scaling in this phase.  We thus have a discontinuous jump in the number of  eigenstates with nonvanishing deviation as we cross the critical point at $\alpha=\frac{1}{2}$.




\begin{figure}[t]
\includegraphics[width = \columnwidth]{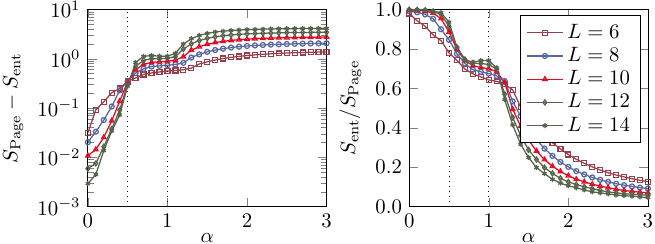}
\caption{Properties of the  $\lceil 0.2 N^\alpha \rceil$ eigenstates with the highest and lowest energies.  Ensemble-averaged eigenstate entanglement entropy $S_\text{ent}$ is compared to the Page value $S_\text{Page}$ in two different ways. The vertical dotted lines mark $\alpha = \frac{1}{2}$ and $\alpha = 1$, bounding the weakly ergodic phase.  Left: For $\frac{1}{2} < \alpha < 1$, the deviation increases with system size $L$, so there is a nonzero deviation from the Page value in the thermodynamic ($L\to\infty$) limit.  Right:  the ratio $S_\text{ent}/S_\text{Page}$ is not decreasing with $L$, so the majority of these eigenstates have volume law scaling.}
\label{fig: critical}
\end{figure}

\begin{figure}[t]
\includegraphics[width = \columnwidth]{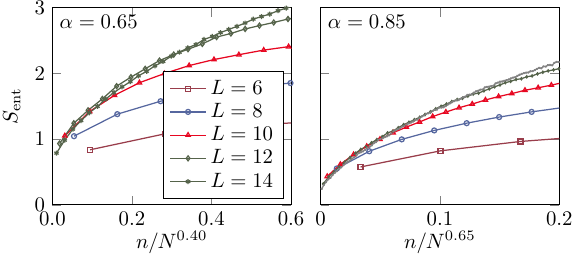}
\caption{Collapse of the $L = 12$ and $L = 14$ curves for $\alpha= 0.65$ with $\delta_2 = 0.40$ (left) and $\alpha = 0.85$ with $\delta_2 = 0.65$ (right) when plotting the ensemble-averaged eigenstate entanglement entropy of the $n$-th eigenstate as a function of $n/N^{\delta_2}$.}
\label{fig: collapse}
\end{figure}

\begin{figure}[t]
\includegraphics[width = \columnwidth]{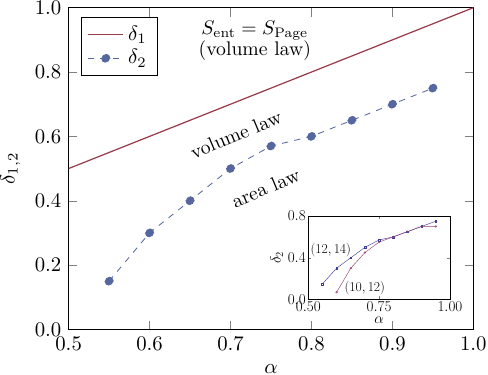}
\caption{Scaling of the entanglement entropy of the $N^\delta$-th eigenstate as a function of $\alpha$. The coefficient $\delta_1$ separates two types of volume-law eigenstates, whereas $\delta_2$ separates volume-law eigenstates from area-law eigenstates for the largest studied system size. The estimates shown for $\delta_2$ are lower bounds. The inset shows the estimates of $\delta_2$ for collapses between the data for system sizes $L$ and $L + 2$.}
\label{fig: delta-alpha}
\end{figure}

We uncover a second boundary between edge and bulk behaviors by plotting $S_\text{ent}$ directly (instead of plotting $S_\text{Page} - S_\text{ent}$) against the scaled eigenstate index $n/N^{\delta_2}$.   If the $S_\text{ent}$ versus $n/N^{\delta_2}$ curves for different sizes coincide for small  $n/N^{\delta_2}$ up to a finite value, this indicates an area law ($S_\text{ent}\sim L^0$) for the lowest $\sim N^{\delta_2}$ eigenstates.  Thus, we look for a collapse of the $S_\text{ent}$ versus $n/N^{\delta_2}$ curves for some value of $\delta_2$.  Figure~\ref{fig: collapse} shows examples of such collapses.   Due to finite-size effects, we can only infer a lower bound for $\delta_2$ from our data, as the inferred value of $\delta_2$ still increases with increasing system size $L$, for the system sizes available to us.   This $L$-dependence is shown in the inset to Fig.~\ref{fig: delta-alpha}. The main panel, in turn, shows the lower bound estimate for $\delta_2$, obtained from the two largest system sizes at our disposal, $(L,L+2)=(12,14)$.  

To synthesize these results, in Fig.~\ref{fig: delta-alpha} we show the $\alpha$-dependence of both $\delta_1$ and $\delta_2$. About $\sim N^{\delta_2}$ eigenstates with lowest and highest eigenenergies are area-law states.  The number of area-law eigenstates appears to grow from $\mathcal{O}(1)$ at $\alpha=0.5$ to an extensive number at $\alpha=1$. We expect that at $\alpha>1$ most eigenstates show area-law entanglement, meaning $\delta_2 = 1$, although from the currently available data we can not explicitly confirm this expectation. 
There is an intermediate set of ${\sim}N^{\delta_1}$ eigenstates which have volume law scaling but a nonvanishing deviation from the Page value.  Finally, all the other eigenstates, a measure one fraction, have proper volume law scaling with $S_\text{ent} \to S_\text{Page}$ in the thermodynamic limit.

\section{Concluding discussion} \label{sec: conclusions}

In this work we explored interpretations of the PLBRM ensemble as Hamiltonians of one-dimensional quantum many-body systems, either disordered or chaotic.  This was motivated by the observation that this class of random matrices, in the weakly ergodic phase $\frac{1}{2} < \alpha < 1$, displays the distinction between edge and bulk eigenstates that is typical of many-body quantum systems.  In many-body systems with local interactions, this spectral edge-bulk distinction leads to the eigenstate entanglement entropy having a characteristic rainbow shape, which is absent in the GOE and GUE ensembles but present in the weakly ergodic phase of the PLBRM ensemble. 

In addition, random matrices are natural models for disordered quantum many-body systems~\cite{Vojta19}, which are intensely studied currently, e.g., in  the context of many-body localization~\cite{Abanin19, Sierant24} and quantum computation~\cite{Gyenis21, Berke22, Boerner24}.  Thus, we expect the many-body interpretation of PLBRMs to be of interest for studies on ergodicity breaking and localization in disordered interacting quantum many-body systems. 
This perspective has been pursued for both the PLBRM ensemble~\cite{Rao22} and for the closely related ultrametric ensemble~\cite{Suntajs_Hopjan_DeRoeck_Vidmar_PRR2024_avalanche_ultrametric}.

We proposed and compared three different ways to assign many-body product basis states to the basis vectors of the random matrices.  Our Gray code labeling scheme improves on the previously used binary labeling scheme.  We explained the need to apply ``site randomization'' to ensure that the resulting many-body Hamiltonians are spatially uniform.  Using this setup, we focused on a particular physical property of the resulting Hamiltonians, namely, the half-system bipartite eigenstate entanglement entropy $S_\text{ent}$.  We 
examined the behavior of $S_\text{ent}$ in each of the phases.  For all the labeling schemes, the common features are: 
(1) In the fully ergodic phase, $\alpha < \frac{1}{2}$, the eigenstate entanglement entropy saturates the bound for fully ergodic states ($S_\text{ent}\to S_\text{Page}$) over the complete spectrum. 
(2) In the weakly ergodic phase, $\frac{1}{2} < \alpha < 1$, there is significant variation across the spectrum: the spectral bulk and spectral edge eigenstates have different scaling behaviors.   
(3) In the localized phase $\alpha>1$, the complete spectrum shows area law or at least sub-volume law behavior.  

For the thermodynamic limit, these behaviors imply, firstly, that a sharp change should be seen in the spectral edge $S_\text{ent}$ at the critical point between the fully and weakly ergodic phases (at $\alpha=\frac{1}{2}$), and secondly, that a sharp change should be seen in the bulk $S_\text{ent}$ at the critical point between the weakly ergodic and localized phases (at $\alpha=1$).  We have confirmed this expectation by analyzing the finite-size crossings in the $S_\text{ent}$  versus $\alpha$ curves.  

The energy-dependence is arguably the most interesting  in the intermediate weakly ergodic phase.  This is also the regime that models the rainbow behavior of $S_\text{ent}$, which was a primary motivation behind this study.  We therefore performed a more detailed study of this regime.  We have managed to demarcate the eigenstates into three types according to their entanglement scaling behaviors.   The spectral bulk eigenstates show entanglement entropy equal to these for maximally ergodic states ($S_\text{ent} \to S_\text{Page}$), the spectral edge eigenstates show area law, and there is an intermediate set of eigenstates that have volume law but a deviation $S_\text{Page}-S_\text{ent}$ that is non-decreasing with system size.   The structure is richer than a single demarcation between the area law eigenstates and the $S_\text{ent}{\to}S_\text{ent}$ eigenstates.   

Of the two exponents marking the two boundaries in the spectrum, $\delta_1$, which separates decreasing from increasing deviations from $S_\text{Page}$, follows the simple behavior $\delta_1=\alpha$.  The other exponent, $\delta_2$, marking the area law eigenstates at the very edges of the spectrum, was numerically more challenging to calculate: our procedure only provides a lower bound.   The data is consistent with $\delta_2$ increasing from $0$ to $1$ as $\alpha$ increases from $\frac{1}{2}$ to $1$, although this behavior near $\alpha=1$ is challenging to confirm (Fig.~\ref{fig: delta-alpha}).  We pointed out that the result $\delta_1=\alpha$ has the following curious consequence.   As one approaches the fully ergodic phase  ($\alpha\to \frac{1}{2}+0^+$), one might naively expect that the number of eigenstates that deviate from $S_\text{Page}$ should vanish.  However, our result $\delta_1=\alpha$ indicates that  $\sim N^{1/2}$  eigenstates show nonvanishing deviation from the Page value, for $\alpha\to \frac{1}{2}+0^+$.  This is a vanishing fraction, but a diverging number.  
This type of transition is similar to the quantum Zeno transition for the inverse participation ratio, $\text{IPR} = c_2 N^{-D_2}$, where the fractal dimension $D_2$, the analog of our critical exponent $\delta_1$, does not undergo any transition, while the prefactor $c_2$ (the analog of our $S_\text{Page} - S_\text{ent}$) develops a jump at the transition from a finite to zero value~\cite{Sierant22}.

Recent analytic work of interest to the present line of inquiry includes studies of entanglement in Gaussian eigenstates~\cite{LiuChenBalents_PRB2018_SYK, Lydzba_Rigol_Vidmar_PRL2020_eigenstateEE_randomquadratic, Lydzba21, Bianchi21, Bianchi_Hackl_Kieburg_Rigol_Vidmar_PRXQ2022}, and a treatment of the crossover between edge and bulk behavior in conformal field theory and free-fermion systems~\cite{MiaoBarthel_PRL2021}. We are not aware of cases where the spectral boundaries between different eigenstate entanglement behaviors have been quantified.

Our work opens up a number of open questions: 

(1) Many-body systems which thermalize have similar entanglement profiles as our weakly ergodic phase.  It would be interesting to try to demarcate area-law edge states from volume-law  bulk states in such a case; we are not aware of a many-body system where such a boundary has been quantitatively identified.  An additional interesting question for many-body systems is: whether there might be an intermediate set of eigenstates with nonvanishing deviation as in the present case, or other type of intermediate behavior. 

(2) Our numerical result $\delta_1=\alpha$ is strikingly simple, suggesting that an analytic derivation of the exponent $\delta_1$ might be possible.  This remains an open task. 

(3) The $\delta_2$-$\alpha$ curve raises more questions.  Since we were restricted to a lower bound and the numerical determination was challenging, a way to determine this dependence with more numerical certainty would be useful. An analytic understanding of this exponent also remains an open question. 

(4) There are several questions of convention-dependence.  We have chosen a particular definition of $\alpha$, through our choice of the $a(r)$ function in Eq.~\eqref{eq:a} defining the PLBRM matrix structure.  We do not expect any of our results to change qualitatively if one of the other standard definitions are used, as long as the bandwidth is $\mathcal{O}(1)$.  However, once the bandwidth $\beta$ is large, one might expect to start incorporating the physics of sharply banded random matrices~\cite{Wigner_1955_banded, Fyodorov_Mirlin_PRL1991_banded}.  It may be a fruitful endeavor to interpolate between the physics of these two ensembles of random matrices. 

(5) Another interesting question of convention is the issue of labeling schemes connecting PLBRM indices to many-body basis states.  For our analysis of the energy-dependence of $S_\text{ent}$ scaling in the weakly ergodic phase (Sec.~\ref{sec: boundary}), we focused on the Gray code scheme.  It is possible that these results might depend on the labeling scheme, e.g., the random or binary schemes might have different demarcation behaviors between spectral edge and bulk, or different values of $\delta_1$ and $\delta_2$.  It would therefore be fruitful in future work to map out the dependence on labeling scheme.  

(6) Through the Gray code scheme and the site randomization procedure, this work provides a basis for using PLBRMs as models for many-body systems.  We have analyzed static Hamiltonians in this framework, but one could extend this line of work to model open or driven quantum many-body systems. 

(7) In this work we focused on the PLBRM class as a minimal model for displaying and analyzing the entanglement rainbow structure.  One can  extend  the GOE/GUE ensembles in various other ways to incorporate features of many-body Hamiltonians, e.g., one could include sparsity, locality etc.

\acknowledgments
MH acknowledges support from the Deutsche Forschungsgemeinschaft under Grant No. SFB 1143 (Project ID No 247310070). IMK acknowledges support by the European Research Council under the European Union’s Seventh Framework Program Synergy ERC-2018-SyG HERO-810451.  MH thanks M.~Rigol and L.~Vidmar for fruitful discussions and feedback.

\bibliography{library-paper}
\end{document}